\documentstyle[preprint,aps]{revtex}
\begin{document}
\draft

\title{Error threshold in finite populations}

\author{D. Alves  and J. F. Fontanari }

\address{
Instituto de F\'{\i}sica de S\~ao Carlos  \\ Universidade de S\~ao
Paulo \\ Caixa Postal 369 \\ 13560-970 S\~ao Carlos SP \\ Brazil}


\maketitle

\begin{abstract}

A simple analytical framework 
to study the molecular  quasispecies evolution of finite populations
is proposed, in which the population is assumed to be 
a random combination of the constituent molecules
in each generation, i.e., linkage disequilibrium at the population
level is neglected. 
In particular, 
for the single-sharp-peak replication landscape
we investigate the dependence of the error threshold on the population 
size  and find that
the replication accuracy at the threshold increases 
linearly with the reciprocal of the population size
for sufficiently large populations.
Furthermore, in the deterministic limit
our formulation  yields the exact steady-state of the quasispecies
model, indicating  then that  the population composition is
a random combination of the molecules.

\end{abstract}



\section{Introduction}

An important  issue in the investigation of  
the dynamics  of competing self-reproducing  macromolecules,
whose paradigm is Eigen's  quasispecies model
\cite{Eigen}, is the effect of the finite size of the 
population  on the error threshold phenomenon
that limits the length of the molecules \cite{reviews}.
The quasispecies model was originally formulated as a deterministic
kinetic theory described by a set of ordinary differential
equations for the concentrations of the different types
of molecules that compose the population. Such  formulation,
however, is valid only in the limit where the total number of
molecules  $N$  goes to infinity.
More pointedly,
in this  model a molecule is represented by a string of $\nu$ digits 
$\left (s_1, s_2, \ldots, s_\nu \right )$, with
the variables $s_\alpha$ allowed to take on $\kappa$ different
values, each of which representing a different type of monomer
used to build the molecule. For sake of simplicity,
in this paper  we will consider only binary strings, i.e., 
$s_\alpha = 0,1$. 
The concentrations $x_i$ of molecules of 
type $i =1, 2,  \ldots, 2^\nu $ evolve in time according to 
the following differential equations \cite{Eigen,reviews}
\begin{equation}\label{ODE}
\frac{dx_i}{dt} = \sum_j W_{ij} x_j - \left [ D_i + \Phi \left ( t
\right ) \right ] x_i \;  ,
\end{equation}
where the constants $D_i$ stand for the death probability of 
molecules of type $i$, and $\Phi (t)$ is a dilution flux 
that keeps the total concentration constant. This flux 
introduces a nonlinearity in  (\ref{ODE}), and is determined
by the condition $ \sum_i dx_i /dt = 0$. 
The elements of the 
replication matrix $W_{ij}$  depend on the replication rate or
fitness
$A_i$ of the
molecules of 
type $i$  as well as on the Hamming distance $d \left ( i,j \right )$
between strings $i$ and $j$. They are given by 
\begin{equation}
W_{ii} = A_i \,  q^\nu
\end{equation}
and 
\begin{equation}
W_{ij} = A_i \,
 q^{\nu - d \left ( i,j \right )} 
 \left ( 1 - q \right )^{d \left ( i,j \right )} ~~~~i \neq j ,
\end{equation}
where $0 \leq q \leq 1$ is
the single-digit replication accuracy, which is assumed to be the same 
for all digits. Henceforth we will set $D_i = 0$ for all $i$.
The quasispecies concept is illustrated more neatly
for the single-sharp-peak replication 
landscape, in which we 
ascribe the replication rate $a > 1$ to  the so-called master string
$\left (1, 1, \ldots, 1 \right )$, and the replication rate $1$ to  the
remaining strings. 
In this context, the parameter $a$ is termed selective advantage
of the master string.
As the error rate $1-q$ increases, two distinct
regimes are observed in the population composition: the 
{\em quasispecies} regime characterized
by the master string and its close neighbors, and the 
{\em uniform} regime where the $2^\nu$ strings appear in the
same proportion. The transition between these regimes takes place
at the error threshold  $1-q_t$, whose value depends
on the parameters $\nu$ and $a$ \cite{Eigen,reviews}. A
genuine thermodynamic order-disorder phase transition
occurs in the limit
$\nu \rightarrow \infty$  only \cite{Lethausser,Tarazona,Galluccio}.
We must note, however, that standard 
statistical mechanics tools developed to study the 
surface equilibrium properties of lattice systems can be
used to investigate the finite $\nu$ case as well 
\cite{Lethausser,Tarazona}. Moreover, the complete
analytical solution of the single-sharp-peak replication 
landscape has been found recently by mapping the stationary solution
of the kinetic equations (\ref{ODE}) into a polymer localization
problem \cite{Galluccio,Gal}.

Closely related  to our approach to the quasispecies
evolution of finite populations  is
the population genetics formulation of the deterministic 
quasispecies model proposed recently \cite{Alves}.
In that formulation  it is assumed that the molecules are
characterized solely by the  number of monomers $1$
they have, regardless of  the particular
positions of these monomers inside the molecules. 
Hence there are only $\nu+1$ different
types of molecules which are labeled by the integer 
$P = 0,1,\ldots, \nu$. 
This assumption is not so far-fetched since the feature that 
distinguishes the molecules is their replication rates $A_i$, 
which in most analyses
have been chosen to depend on $P$ only, i.e., $A_i = A_P$ 
\cite{reviews}. Furthermore, denoting the frequency of monomers $1$ 
in generation $t$ by $p_t$, it is assumed that
the molecule frequencies $\Pi_P (t)$
are given by the binomial distribution
\begin{equation}\label{binomial}
      \Pi_P(t)  = \left ( \! \! \begin{array}{c} \nu \\ P \end{array}
      \! \! \right ) \,
      \left ( p_t \right )^{P} \, \left  (1 - p_t \right )^{\nu -P}
\end{equation}
for $P = 0, 1, \ldots, \nu$.
Thus, in each generation  the monomers are sampled with replacement
from a pool containing monomers $1$ and $0$ in the proportions
$p_t$ and $1-p_t$, respectively. This amounts to neglecting linkage
disequilibrium, i.e., in each generation
the molecule frequencies are random combinations
of the constituent monomers \cite{Feldman}.
With the two  assumptions presented above
a simple recursion relation for the  monomer frequency $p_t$ can be
readily derived \cite{Alves}.

To take into account the effect of finite $N$, the deterministic 
kinetic formulation must be  replaced by a stochastic formulation
based on a master equation for
the  probability distribution  of the number of  different types
of molecules in the population \cite{Ebeling,McCaskill}.
However, the extreme  approximations used to derive
results from that master equation or from related Langevin equations
\cite{Inagaki,Yi} have hindered the analysis of
the error threshold for finite populations. 
An alternative   approach to
study  stochastic chemical reaction networks is the
algorithm proposed by Gillespie \cite{Gillespie}, which
has been successfully employed to simulate numerically 
the quasispecies model,  providing
thus a base line for analytical investigations \cite{Nowak}.
The goal of this work is to propose an analytical framework 
to investigate the quasispecies evolution of finite populations.
More specifically, we will focus on the evolution of the molecule 
frequencies $\Pi_P (t)$ for  $P=0,\ldots,\nu$ and, since for finite
$N$ these frequencies are random variables,
we will derive a recursion relation
for the average values  ${\overline \Pi_P} (t)$.
Although we will concentrate mainly  on the dependence 
of the error threshold on the population size $N$, the formalism
presented in the sequel can be applied to study
a variety
of fascinating phenomena related to the finitude of the population,
such as
mutational meltdown \cite{Gabriel} and punctuated equilibria or stasis
\cite{Yi,Crutch}, to mention only a few. Moreover, since modern theories
of integration of information in pre-biotic systems involve
the compartmentation of a small number of molecules (typically 10 to
100) \cite{book}, the understanding of the effects of the error 
propagation in finite populations has become an important
issue to the theories of the origin of life.

\section{The model}

In each generation the population is described by the  vector 
$ {\bf{n}} = \left ( n_0, \ldots, n_\nu \right )$
where $n_P$ is the number of molecules of type
$P$, so that  $\sum_P n_P = N$. 
Similarly to the deterministic case \cite{Alves}, we have to resort to 
a simplifying  assumption to relate the molecule
frequencies $\Pi_P$ to the  vector ${\bf n}$. In
particular, in generation $t$
we consider a molecule pool containing 
the different molecule types
in the proportions $\Pi_P $, so that ${\bf{n}}$ is distributed by the 
multinomial distribution
\begin{equation}\label{P_n}
{\cal{P}} \left ( {\bf{n}} \right ) = 
\frac{N!}{n_0! \, n_1! \ldots n_\nu!}
\, \left [ \Pi_0 (t) \right ]^{n_0} \left [ \Pi_1 (t) \right]^{n_1} 
\ldots  \left [ \Pi_\nu (t) \right ]^{n_\nu} .
\end{equation}
Hence in each generation the molecules are sampled with replacement 
from the molecule pool. In this sense, in each generation
the  population is a  random combination
of the constituent molecules, which 
amounts to neglecting linkage
disequilibrium at the population level. More pointedly,
the population composed of the offspring 
of the  molecules present
in the generation $t-1$ is destroyed and its molecule 
frequency $\Pi_P (t)$
used to create an  entire new population according to (\ref{P_n}).
Although
this  procedure destroys the correlations between the molecules,
it does not cause any significant loss of genetic information since 
the fitness of the molecules 
depend only on the number of monomers $1$ they have, which, in the
average, is not affected by the procedure. 

The changes in the population composition ${\bf n}$  
are due to the driving of
natural selection, modeled by the replication rate $A_P$, and to
mutations, modeled by the error rate per digit $1-q$. 
Following the 
prescription used in the implementation of  the standard
genetic algorithm \cite{Goldberg}, 
we consider
first the effect of natural selection and then the effect of
mutations.
As usual we assume that the number of offspring
that each molecule contributes to the new generation 
is proportional to its relative replication rate which, for molecules 
of type
$P$, is defined by  
\begin{equation}
W_P \left ( {\bf n} \right ) = \frac{ n_P A_P }{\sum_R n_R A_R} .
\end{equation}
Thus the population composition  after selection 
is described by the random vector
${\bf n}'= \left ( n_0', \ldots, n_\nu' \right )$ 
which is distributed according to
the conditional probability distribution 
\begin{equation}\label{P_s}
{\cal{P}}_s \left ( {\bf{n}}' \mid {\bf{n}} \right ) = 
\frac{N!}{n_0'! \, n_1'!  \ldots n_\nu'!}
\, \left [ W_0 \left ( {\bf n} \right ) \right ]^{n_0'} 
\left [ W_1 \left ( {\bf n} \right ) \right ]^{n_1'} 
\ldots \left [ W_\nu \left ( {\bf n} \right ) \right ]^{n_\nu'} .
\end{equation}
Next we consider the changes in ${\bf n}'$ due to mutations. 
After mutation, the population is described by 
${\bf n}'' = \left ( n_0'', \ldots, n_\nu'' \right )$ 
whose components are written as
\begin{equation}
n_P'' = \sum_{R=0}^\nu n_{PR}'' ,
\end{equation}
where the integer  $n_{PR}''$ stands for the number of 
molecules of type
$R$ that have mutated to a molecule of type $P$. 
Clearly, $n_R' = \sum_P n_{PR}''$.
We note that the
probability of mutation from a molecule of type $R$ to a
molecule of type $P$ is given by 
\begin{equation}\label{M}
M_{P R} = \sum_{Q = Q_l}^{Q_u} \left ( \! \! \begin{array}{c} R \\ Q 
\end{array}
\! \! \right ) \, \left ( \! \! \begin{array}{c} \nu - R \\ P - Q 
\end{array} \! \! \right ) \, q^{\nu - P - R + 2Q} \,
\left ( 1 - q \right )^{P + R - 2Q} ,
\end{equation}
where $Q_l = \mbox{max} \left ( 0,P+R-\nu \right)$ and
$Q_u = \mbox{min} \left ( P,R  \right)$.
The population
is more conveniently described  by the set 
 $\{ n_{PR}'' \}$ rather than by ${\bf n}''$. In fact,
given $ n_R'$ the conditional  probability  distribution
of $\{ n_{PR}''\}$
is again a multinomial
\begin{equation}
{\cal{P}}_m \left (n_{0R}'',n_{1R}'',\ldots, n_{\nu R}'' 
\mid n_R' \right ) = \frac{n_R' !}{n_{0R}''! \, n_{1R}''!
\ldots n_{\nu R}'' !} \, M_{0R}^{n_{0R}''} \, M_{1R}^{n_{1R}''} \ldots
M_{\nu R}^{n_{\nu R}''}
\end{equation}
for  $R=0,\ldots, \nu$. 
In this framework the frequency of molecules of type $P$ 
in the next generation $\Pi_P ( t+1 )$ is given simply by 
$ \frac{1}{N} \sum_R  n_{PR}''$. This
frequency is used to generate the new population of $N$ molecules 
of length $\nu$ according to the distribution
(\ref{P_n}). The procedure is then repeated again.

We have run simulations for  the single-sharp-peak replication landscape
using the procedure described above, which neglects linkage disequilibrium
at the population level, as well as the standard genetic algorithm
\cite{Goldberg}, in which the correlations between consecutive
generations are maintained. We have focused on the
effect of the error rate $1-q$ on the normalized
mean Hamming distance $d$ between the master string
and the whole population in the stationary regime.
This quantity is given by
the fraction of monomers $0$ in the entire  population, i.e., 
$d = \frac{1}{\nu} \sum_P (\nu - P) \Pi_P $.
In Figs.\ (\ref{f1}) and (\ref{f2})  we present the results of
the simulations for 
$d$ and its standard deviation $\sigma$, respectively,
as functions of the error rate $1-q$.  
The initial population is set with $\Pi_\nu = 1$ and
$\Pi_P = 0 $ for $P \neq \nu$,
and it is left to evolve for $2~ 10^3$ generations. No significant
differences were found for longer runs or for different choices
of the initial molecular frequencies.
Each data point involves two kinds of average: for each 
run we average over the mean Hamming distance in the  
last $100$ generations; this value is then averaged over
$200$ runs.  We note that even if the populations are 
identical in the initial generation, the random character
of the transitions ${\bf n} \rightarrow {\bf n}'
\rightarrow {\bf n}''$ will make them distinct  in the
next generation. It is clear from these results that, as $N$
increases, the quantitative effects of assumption (\ref{P_n}) 
become less significant. Moreover, the dependence of $d$ and
$\sigma$ on the error rate is qualitatively the same for both algorithms. 

\section{Recursion equations}

To derive an  analytical recursion relation for 
the average molecular frequencies $\overline{\Pi}_P (t)$
we consider the following approximate procedure, 
akin to the annealed approximation of the 
statistical mechanics of disordered systems,
which facilitates greatly 
the analysis: instead of averaging
over the populations only after the stationary regime is reached,
we perform this average in each generation. The result obtained
$ \overline{\Pi}_P (t) $ is then used to build the new populations. 
Of course, in doing so we neglect the fluctuations of $\Pi_P (t)$ 
for the different runs.  
Within this framework the average frequency of molecules of type
$P$  in
generation $t+1$ is written as
\begin{equation}\label{av_1}
\overline{\Pi}_P (t+1)  = \frac{1}{N } \, 
 \sum_{{\bf n}} \sum_{{\bf n}'}
\sum_{\{n_{PR}''\}}  \sum_R   \, n_{PR}'' \,
{\cal{P}}_m \left( \{n_{PR}''\} \mid {\bf n}'\right)
{\cal{P}}_s \left( {\bf n}' \mid {\bf n} \right)
{\cal{P}} \left ( {\bf n} \right ) .
\end{equation}
Using
\begin{equation}
\sum_{\{ n_{PR}'' \}} \, n_{PR}'' \,
{\cal{P}}_m \left( \{n_{PR}''\} \mid {\bf n}'\right) =
M_{PR} n_R'
\end{equation}
and 
\begin{equation}
\sum_{n_R'}  n_R' {\cal{P}}_s \left( {\bf n}' \mid {\bf n} \right)
= N W_R  \left ( {\bf n} \right )
\end{equation}
we rewrite (\ref{av_1}) as
\begin{equation}\label{av_2}
\overline{\Pi}_P (t+1)  =  \sum_{{\bf n}}
\sum_R \, M_{PR} W_R  \left ( {\bf n} \right ) 
{\cal{P}} \left ( {\bf n} \right ) .
\end{equation}
Noting that 
 $\sum_P M_{PR} = 1$ and $\sum_R W_R \left ( {\bf n} \right )
=1 $, we can  easily verify that the normalization condition
$\sum_P \overline{\Pi}_P (t+1) = 1$ is satisfied.
To proceed further we must specify the replication rate $A_P$.
In the case of the single-sharp-peak replication landscape, i.e.,
$A_\nu = a$ and $A_P = 1$ for $P \neq \nu$,
the summations over
$n_0, \ldots,n_{\nu-1}$ can be
readily carried out. The final result is
\begin{equation}\label{rec}
\overline{\Pi}_P (t+1)  = 
M_{P \nu} \left [ \overline{\Pi}_\nu (t) \right ]^N
+ \sum_{n=0}^{N-1} B_n \frac{ \sum_{R =0}^{\nu - 1} 
\overline{\Pi}_R (t) \left [ M_{PR} + a  \frac{r}{1-r} M_{P \nu} 
\right ]}
{ 1 + r \left ( a -1 \right )} 
\end{equation}
for $P = 0, \ldots, \nu $.
Here
\begin{equation}
B_n = \left ( \! \! \begin{array}{c}
N-1 \\ n \end{array} \! \! \right ) \, 
\left [ \overline{\Pi}_\nu (t) \right ]^n 
\left [ 1 - \overline{\Pi}_\nu (t) \right ]^{N-1-n} ,
\end{equation}
and $ r = n/N$. Thus, given the initial average molecular frequencies 
$\overline{\Pi}_P (t=0)$ for $P = 0, \ldots, \nu$, equations
(\ref{rec}) are iterated till the stationary regime is reached.

Before we proceed  on the analysis of the stationary solutions
of the recursion equations (\ref{rec}),
some comments regarding the definition of the error threshold are in
order.  
A  popular definition of error threshold
is  the error rate  at which the master
frequency $\Pi_\nu$ vanishes \cite{Eigen,reviews}. The problem with
this definition is that, even in
the deterministic limit $N \rightarrow \infty$, 
$\Pi_\nu$ never vanishes for {\it finite} $\nu$. The vanishing of the
master frequency is an artifact of neglecting 
reverse mutations \cite{Eigen}, which can be justified in the limit  
$\nu \rightarrow \infty$ only.
In particular, for the single-sharp-peak replication 
landscape
this prescription yields a very simple equation for the replication
accuracy at the threshold in the deterministic regime \cite{Eigen,reviews},
\begin{equation}\label{simple}
 - \ln q_t = \frac{1}{\nu} \, \ln a .
\end{equation}
We must emphasize that for finite $\nu$ this equation is
an approximation only. A more appropriate
definition of the error threshold, which is useful
for the finite $N$ case as well, is obtained by considering the statistical
properties of the entire molecular population  \cite{Wiehe}. 
In particular, we focus on the normalized mean Hamming distance $d$
between the master sequence and the entire population and define 
the error threshold
as the error rate at which the standard deviation $\sigma$
is maximal \cite{Wiehe}.

In Figs.\ \ref{f1} and \ref{f2} we show the theoretical predictions
for $d$ and $\sigma$ using the steady-state solution
of the recursion equations (\ref{rec}). As expected, the effects of the 
fluctuations in $\Pi_P (t)$  for the different runs are stronger
for small $N$ and hence our analytical approximation yields very poor
results in this case, although it reproduces quite well the qualitative
behavior pattern of the quantities measured. However, already for 
$N=100$ there is a good agreement 
between the theoretical predictions for $d$ and the simulation
results, provided
that $1-q$  is not too near the threshold transition. Rather
surprisingly, that agreement is better for the 
standard genetic algorithm. 
As expected, however, the theoretical predictions for the standard
deviation $\sigma$ are very poor since our approximation scheme neglects
the fluctuations in $\Pi_P$ between the different runs, which
are directly measured by $\sigma$. Nevertheless, the qualitative 
features
of the simulation results  are again well described by the theoretical
curves. We note, in particular, the abrupt increase of $\sigma$ as 
the error threshold is approached from below and the slow decay as 
the error rate increases further.  
The agreement between theory and simulation
becomes better as $N$ increases. 
In Fig.\ \ref{f3} we show the replication accuracy
at the threshold $q_t$ as a function of   the reciprocal 
of the population size. 
The  increase of $q_t$ with decreasing $N$ is expected since
the fluctuations become stronger for small $N$ and so the 
replication must be more accurate in order to keep the
master string in the population. In particular, 
for large $N$ we find that $q_t$ increases linearly
with $1/N$.  This result is in  disagreement with the predictions 
of the birth and
death model of error threshold proposed by Nowak and Schuster,
which predicts that $q_t$ increases with $1/\sqrt{N}$ for large
$N$ \cite{Nowak}. We note that, despite the claim of those authors, it is 
not possible to discern whether $q_t$ increases with $1/N$ or $1/\sqrt{N}$ 
from their numerical data obtained using Gillespie's algorithm \cite{Nowak}. 

Another interesting phenomenon, termed stochastic escape, 
is the loss of the master string in a finite population
\cite{Higgs1,Higgs,Peliti}. In the limit  $\nu \rightarrow \infty$ 
this loss becomes irreversible, since no reverse mutation will 
be able to restore the master string. We can
easily derive a lower bound to the probability that the master
string is absent from the population using the inequality
\begin{equation}
1 - \bar{n}_\nu \leq Pr \{ n_\nu  = 0 \} ,
\end{equation}
which follows trivially from the fact that $n_\nu \geq 0$.
Using $\bar{n}_\nu = N \bar{\Pi}_\nu$, we can find the replication 
accuracy
$q_{l}$ such that the condition $\bar{\Pi}_\nu = 1/N$ is satisfied
for fixed $\nu$ and $a$.
Clearly, for $q < q_l$ the probability that the master 
string is absent from the population is nonzero. However, since this 
probability may be nonzero for $q > q_l$ as well, $q_l$ gives only
a lower bound to the value of the replication accuracy
below which the stochastic escape phenomenon actually takes place. 
This bound
is presented in Fig.\ \ref{f3} as a function of the reciprocal of
the population size. In the limit $N \rightarrow \infty$
we find $q_l \rightarrow 0$ for finite $\nu$, since in the deterministic
regime
$\bar{\Pi}_\nu$  is bounded  by $1/2^{\nu} > 0$.
It is interesting that for $N$ not too large we find $q_l > q_t$ 
so that the master string is likely to be absent from the
quasispecies for replication accuracies in the range $q_t < q < q_l$.

We turn now to the analysis of 
the deterministic regime,
$N \rightarrow \infty$. 
In this case, the sum in Eq.\ (\ref{rec})
is dominated by the closest integer to
$ (N-1)  \overline{\Pi}_\nu (t) $, so that 
$ r \rightarrow \overline{\Pi}_\nu (t) $ 
and  the recursion equations (\ref{rec}) reduce to   
\begin{equation}\label{det}
\overline{\Pi}_P (t+1)  = \frac{ \sum_{R =0}^{\nu - 1} 
M_{PR} \overline{\Pi}_R (t)   + a M_{P \nu} \overline{\Pi}_\nu (t)}
{ 1 + \overline{\Pi}_\nu (t) \left ( a -1 \right )},
\end{equation}
for $P = 0, \ldots, L$.

To appreciate the relevance of the present formulation of the
quasispecies model, we compare it with the exact solution of
the kinetic equations (\ref{ODE}) for finite $\nu$. In fact,
as pointed out by Swetina and Schuster \cite{Swetina},
for the type of replication
landscape considered in this paper,
the $2^\nu$ molecular concentrations $x_i$ can also be grouped into
$\nu +1$ distinct classes according to the number of monomers 1
that compose the molecules. This procedure allows the  description of
the chemical kinetics by only $\nu + 1$ coupled first-order
differential equations. In particular, for the
single-sharp peak landscape the concentrations 
of molecules in class $P = 0,\ldots,\nu$, denoted by $Y_P$ with
$\sum_P Y_P = 1$, obey
the differential equations \cite{Swetina}
\begin{equation}\label{sw}
\frac{dY_P}{dt} = \sum_{R=0}^{\nu -1} M_{PR} \, Y_R  + a Y_\nu M_{P \nu}
- Y_\nu \left [ 1 + Y_\nu \left ( a - 1 \right ) \right ] .  
\end{equation}
It is clear then that both models (\ref{det}) and (\ref{sw})
possess the same  stationary state. This very interesting finding 
indicates that in the quasispecies
model there is no linkage disequilibrium at the population
level in the stationary regime, i.e., the  population is 
a random combination of the constituent
molecules. In fact, this results holds true for any
choice of the replication landscape $A_P$, as can be easily verified
by taking the limit $N \rightarrow \infty$ in  Eq.\  (\ref{av_2}).

It is also interesting to compare the deterministic
limit of our model (\ref{det})  with the population 
genetics approach to the deterministic quasispecies model \cite{Alves}. 
This can easily be done by writing down 
a recursion equation for the frequency of monomers 1 in the
population 
$\bar{p}_t = \frac{1}{\nu} \sum_P P \, \overline{\Pi}_P (t)$,
 namely,
\begin{equation}\label{p_d}
\bar{p}_{t+1}  = 1 - q + \left ( 2 q-1 \right )
\, \frac{ \bar{p}_t + \left ( a - 1 \right ) \overline{\Pi}_\nu (t) }
{1 + \left ( a - 1 \right ) \overline{\Pi}_\nu (t) } .
\end{equation}
Clearly, this equation is useless 
since one must
solve the general recursion equations (\ref{det}) in order
to find $ \overline{\Pi}_\nu (t)$. However, the population 
genetics formulation makes use of the binomial assumption 
(\ref{binomial}) to set $ \overline{\Pi}_\nu (t) = \bar{p}_t^\nu$ 
so that the  recursion equation (\ref{p_d}) will involve 
the monomer frequencies only \cite{Alves}. 

In  Fig.\ \ref{f4} we present the logarithm of the replication 
accuracy at 
the threshold $\ln q_t$  as a function of the logarithm of the
selective advantage $\ln a$ for several values of $\nu$ for
the deterministic case. There is
a good  agreement with (\ref{simple}) for $a \approx 1$ and,
as expected, this agreement  becomes
better as $\nu$ increases. We have
also  verified that the prediction of the population genetics 
approach \cite{Alves} yields a very poor approximation 
for the location of the error threshold. Furthermore,
we have verified that $\overline{\Pi}_P$ departs significantly 
from a binomial
distribution only near the threshold transition. Aside this region,
the population genetics approach provides a reliable and concise
description of the deterministic quasispecies model.       

\section{Conclusion}

In this paper we have proposed a simple analytical model,
based on the neglect of linkage disequilibrium, to study
the error propagation  in the quasispecies evolution of finite 
populations.
In particular, our finding  that  in the deterministic regime this model 
yields exactly the same stationary state of the
original kinetics model \cite{Swetina} 
implies that the steady-state molecular population 
of the deterministic quasispecies model is a random assembly
of the component molecules.

Some comments regarding the comparison of our approach with
previous population theoretical analyses of the finite $N$ 
quasispecies model \cite{Higgs,Adam} are in order. 
These works provide approximate formalisms
to study the  evolution of finite populations on a multiplicative
single-peak fitness landscape  without neglecting
linkage disequilibrium. Interestingly, for this fitness landscape, 
which is given by
$A_P = \left ( 1 - \hat{\sigma}  \right )^{\nu - P} $ with 
$0 < \hat{\sigma} < 1$,
the binomial assumption (\ref{binomial}) yields the exact
solution to the deterministic equations for $\Pi_P$ \cite{Higgs}. 
Moreover, in the  weak selection limit ($\hat{\sigma} \ll 1$) 
the $1/N$ corrections to the deterministic value of the
mean Hamming distance between the master sequence and the whole 
population 
can be calculated analytically \cite{Higgs}. An alternative
formalism concentrates on the evolution of the ensemble
average of the first cumulants of the distribution of
fitness in the population \cite{Adam}. It is not clear,
however, whether these quantities can be related to the more natural 
measures of the population composition, namely, $d$  and $\sigma$. 
We note that the location of the error threshold is not
addressed  in these works.

An important open question, which can be answered through 
intensive numerical simulations only, is the dependence of
the replication accuracy at the threshold on the population size  
for large populations. In fact, the numerical data existent in the 
literature do not allow to distinguish  between the $1/N$ dependence 
predicted by our model and the $1/\sqrt{N}$ dependence predicted
by the birth and dead model
\cite{Nowak}.  We think that simulations based on genetic
algorithms rather than  on Gillespie's algorithm may prove more 
effective to address this issue.

\acknowledgments
This work was supported in part by Conselho Nacional de 
Desenvolvimento Cient\'{\i}fico e Tecnol\'ogico (CNPq).

\begin{figure}
\caption{Steady-state normalized mean Hamming distance between  the
master sequence and the whole population as
a function of the error rate per digit for $N = 10$ 
($\bigtriangledown$),
and $N=100$ ($\bigcirc$). The full symbols are the results obtained 
with the 
algorithm that neglects linkage disequilibrium, while the
empty symbols are the results obtained with the standard
genetic algorithm. The 
theoretical prediction is given by the solid curves. The dashed
line is the prediction for $N \rightarrow \infty$.
The parameters are $\nu = 10$ and $a=10$.\label{f1}}
\end{figure}

\begin{figure}
\caption{Steady-state standard deviation of the 
normalized mean Hamming distance between the
master sequence and the whole population as
a function of the error rate per digit.
The parameters and convention are the same as for Fig.\ (\ref{f1}).
  \label{f2}}
\end{figure}

\begin{figure}
\caption{Replication accuracy at the error threshold $q_t$
(solid curves) and lower bound to the replication accuracy below which
the stochastic escape phenomenon occurs $q_l$ (dashed curves)  
as  functions of  the reciprocal of the population
size for $a = 10$ and (from bottom to top) $\nu = 6, 8, \ldots, 20$.
\label{f3}}
\end{figure}

\begin{figure}
\caption{ Logarithm of the replication accuracy at 
the threshold in the deterministic regime as a function of 
the logarithm of the
selective advantage for (from top to bottom ) $\nu = 5,10,15,20 $
and $25$. 
The solid curves are obtained  using Eq.\ (\ref{det}) and the
 dashed straight lines are given by Eq.\ (\ref{simple}). \label{f4}}
\end{figure}

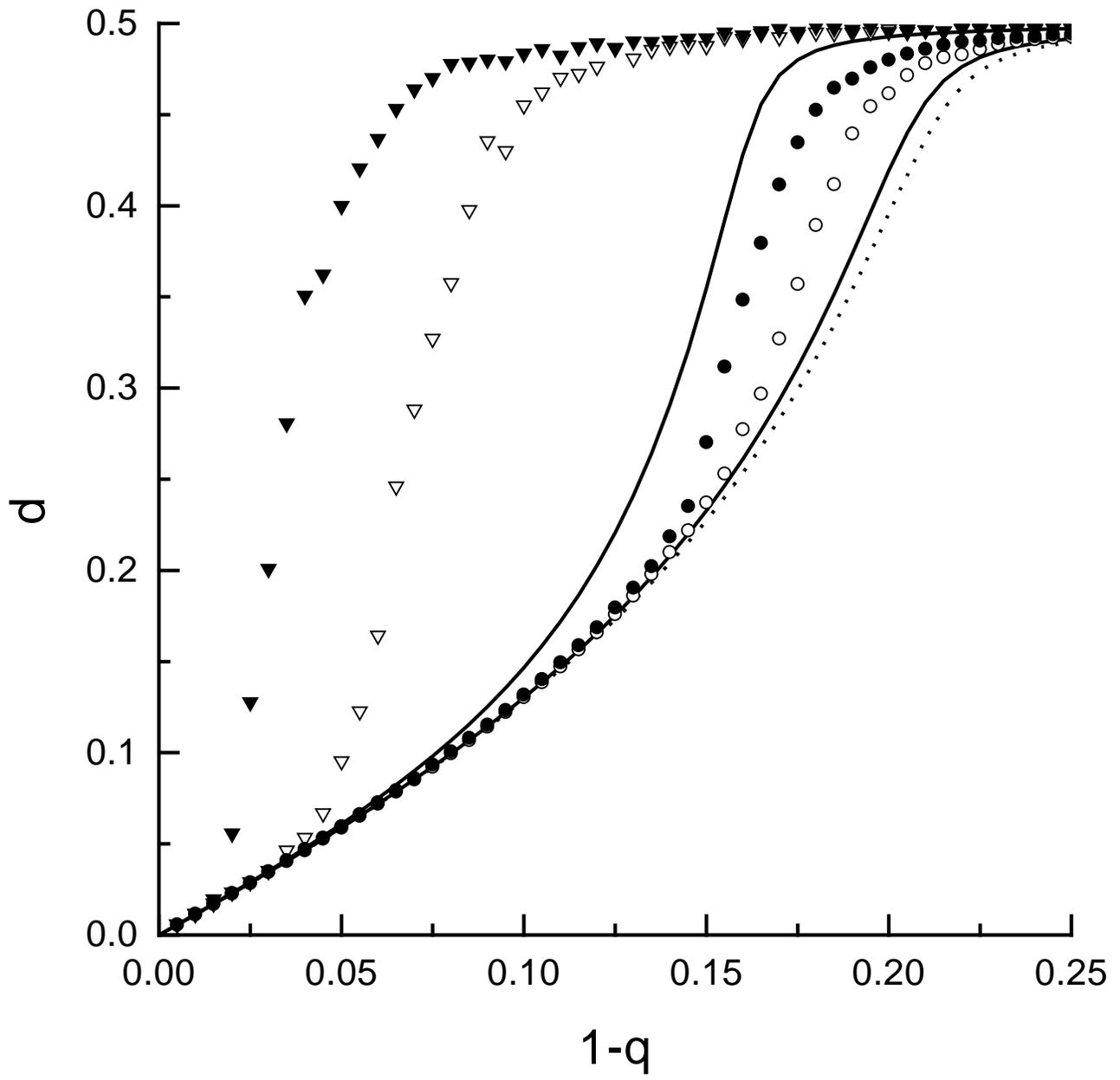
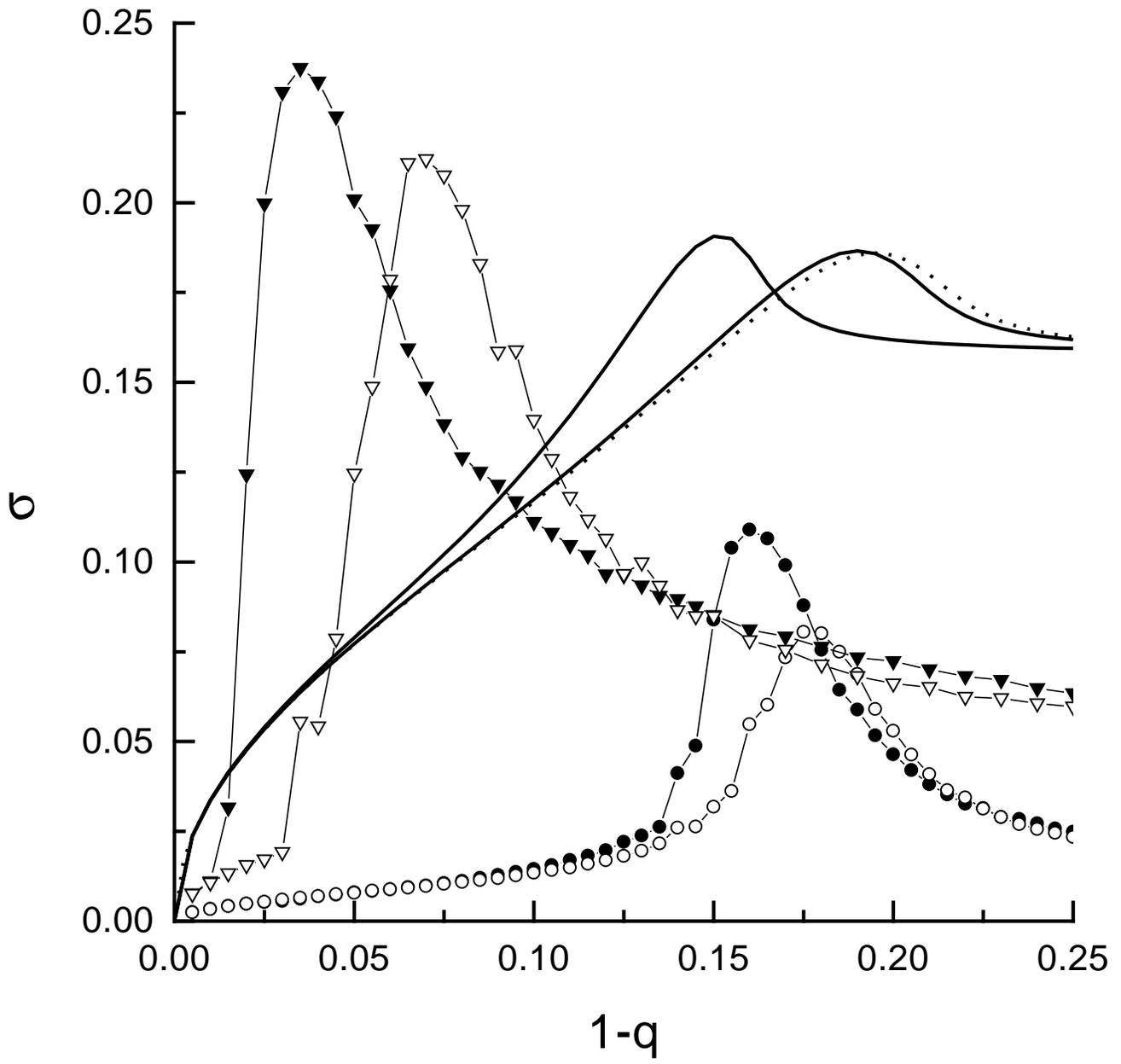
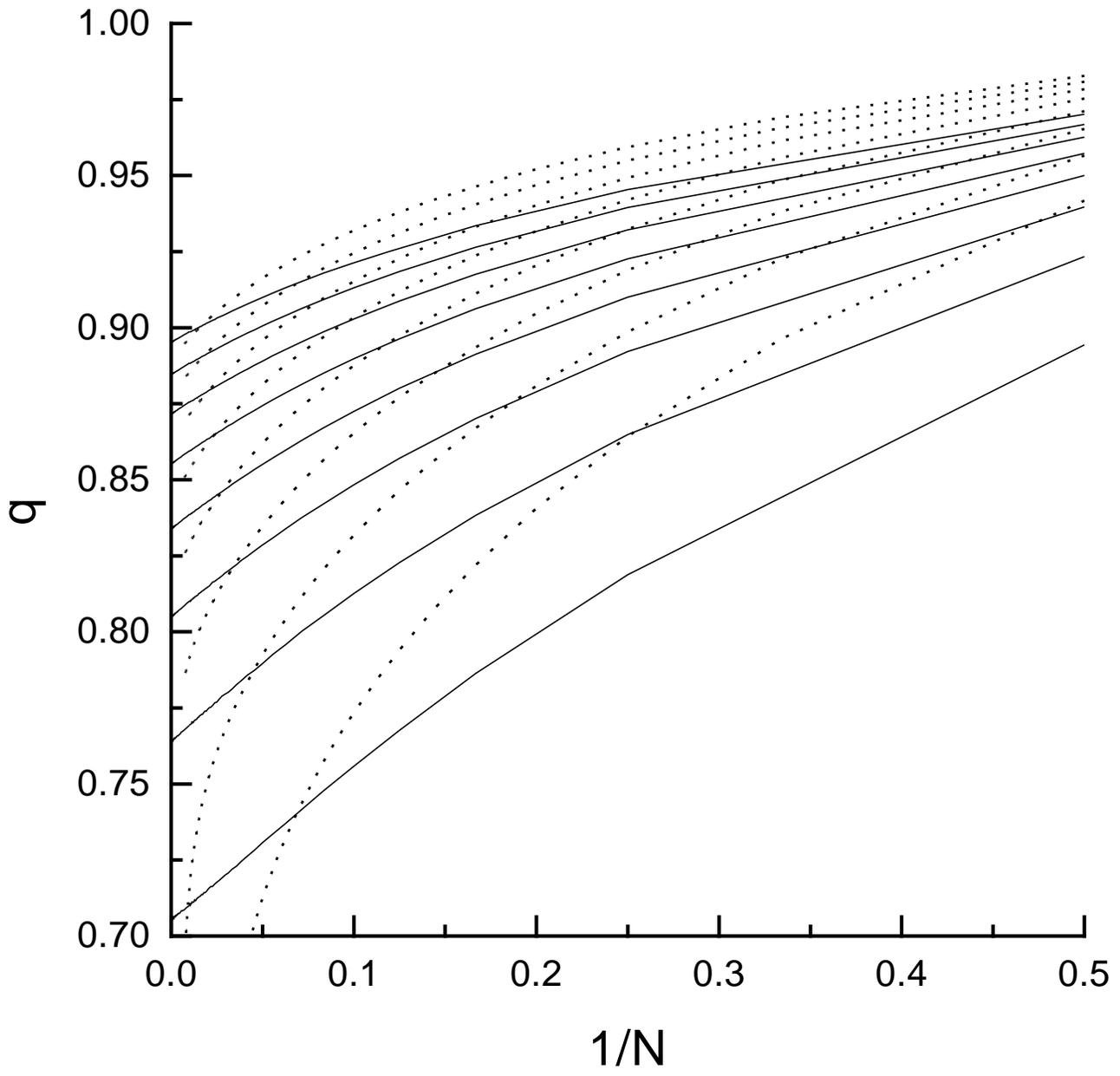
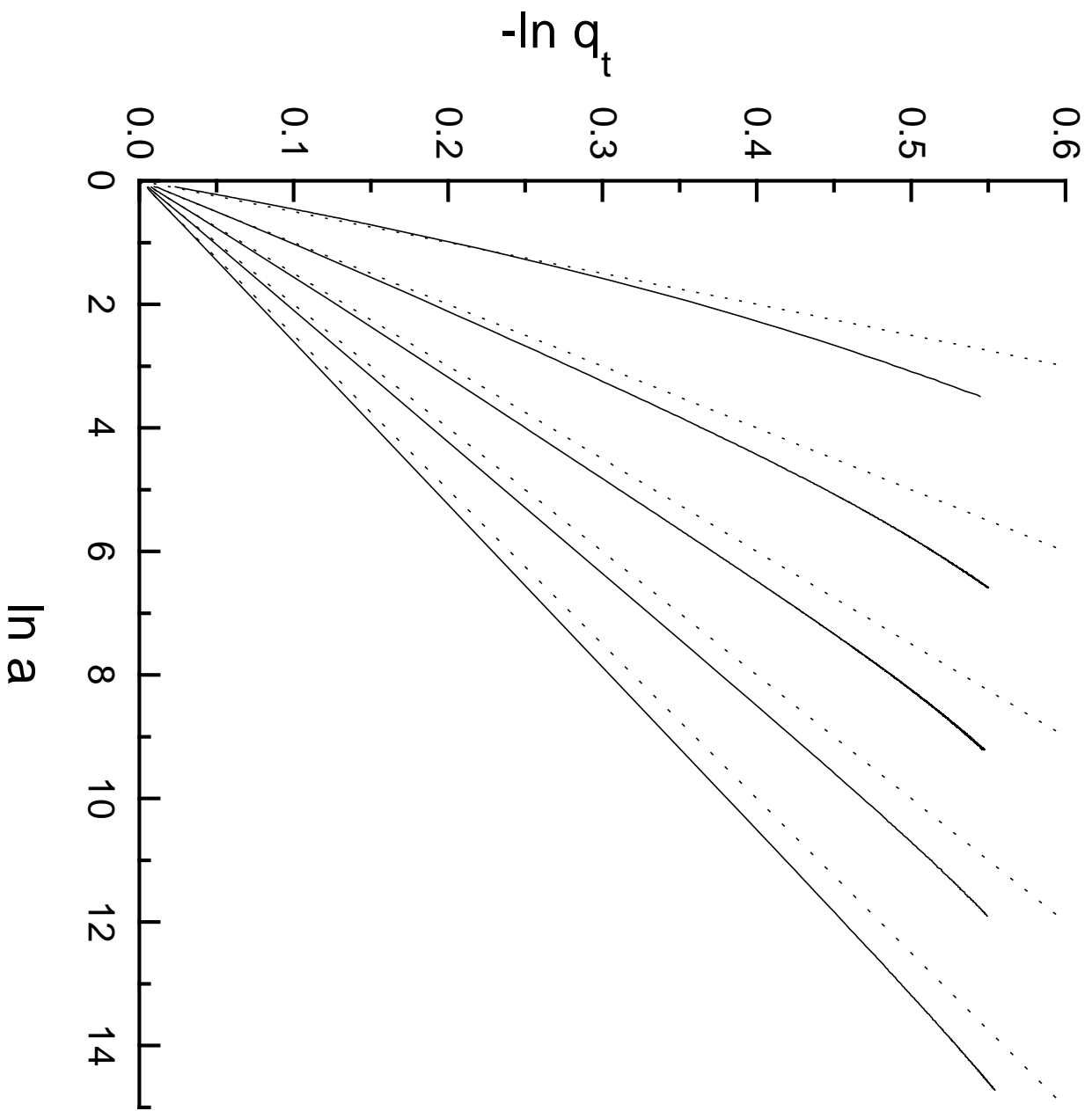
\end{document}